\documentclass[journal,10pt]{IEEEtran}
\IEEEoverridecommandlockouts
\usepackage{cite}
\usepackage{indentfirst}
\usepackage{graphicx}
\usepackage{changepage}
\usepackage{stfloats}
\usepackage{amsfonts,amssymb}
\usepackage{bm}
\usepackage{algorithm}
\usepackage{algorithmic}
\usepackage{amsmath}
\usepackage{amsmath,amssymb,amsfonts}
\usepackage{times}
\usepackage{url}
\usepackage{cite}
\usepackage{textcomp}
\usepackage{xcolor}
\usepackage{color}
\usepackage{setspace}
\usepackage{enumitem}
\usepackage{float}
\usepackage{diagbox}
\usepackage[T1]{fontenc}
\usepackage[utf8]{inputenc}
\usepackage{authblk}
\usepackage{threeparttable}
\usepackage{booktabs}
\usepackage{footnote}
\usepackage{graphicx}
\usepackage{pifont}
\usepackage{subfigure}
\usepackage{mathrsfs}
\newtheorem{pos}{Proposition}
\newenvironment{proof}{{\indent \indent \it Proof:}}{\hfill $\blacksquare$\par}

\def\BibTeX{{\rm B\kern-.05em{\sc i\kern-.025em b}\kern-.08em
    T\kern-.1667em\lower.7ex\hbox{E}\kern-.125emX}}

\renewcommand{\theenumi}{\alph{enumi}}

\allowdisplaybreaks [4]
\begin{document}

\title{Spatial Multiplexing Oriented Channel Reconfiguration in Multi-IRS Aided MIMO Systems}

\author{\IEEEauthorblockN{Yuxuan~Chen, Qingqing~Wu, \textit{Senior Member, IEEE}, Guangji~Chen, Wen~Chen, \textit{Senior Member, IEEE}}
\thanks{ Yuxuan~Chen, Qingqing~Wu, and Wen~Chen are with the Department of Electronic Engineering, Shanghai Jiao Tong University, Shanghai 200240, China (e-mail: {yuxuanchen@sjtu.edu.cn; qingqingwu@sjtu.edu.cn; wenchen@sjtu.edu.cn}). Guangji Chen is with Nanjing University of Science and Technology, Nanjing 210094, China (e-mail:
{guangjichen@njust.edu.cn}).
}
}

\maketitle

\begin{abstract}
Spatial multiplexing plays a significant role in improving the capacity of multiple-input multiple-output (MIMO) communication systems. To improve the spectral efficiency (SE) of a point-to-point MIMO system, we exploit the channel reconfiguration capabilities provided by multiple intelligent reflecting surfaces (IRSs) to enhance the spatial multiplexing. Unlike most existing works, we address both the issues of the IRSs placement and elements allocation. To this end, we first introduce an orthogonal placement strategy to mitigate channel correlation, thereby enabling interference-free multi-stream transmission. Subsequently, we propose a successive convex approximation (SCA)-based approach to jointly optimize the IRS elements and power allocation. Our theoretical analysis unveils that equal IRS elements/power allocation scheme becomes asymptotically optimal as the number of IRS elements and transmit power tend to be infinite. Numerical results demonstrate that when the total number of IRS elements or the power exceeds a certain threshold, a multi-IRS assisted system outperforms a single IRS configuration. 

\end{abstract}
% still approximation


% }
\section{Introduction}

Over the past decade, multiple-input multiple-output (MIMO) systems have significantly enhanced network throughput by simultaneously transmitting multi-stream via the spatial domain. Particularly, the  capacity of MIMO systems increases linearly with the number of sub-channels, which is mainly determined by the rank of the channel matrix\cite{goldsmith}. However, the number of available sub-channels is constrained by the number of independent propagation paths and antennas. 

Recently, intelligent reflecting surface (IRS) has emerged as a promising technology for future wireless systems due to its ability to create customizable propagation environments with lower hardware cost and power consumption compared to traditional antenna arrays \cite{wu2021intelligent,Guangjicite,10159991}. Additionally, IRS can be densely deployed to facilitate data transmission. In addition to passive beamforming, the optimization of IRS deployment provides another degree of freedom for realizing channel customization. For a point-to-point link, the seminal work\cite{Qingqing-2021} unveiled that the passive IRS should be deployed near the transmitter/receiver (Tx/Rx) to minimize the cascaded channel path loss. While the above work focused on applying the deployment of an IRS to increase the received power, it is also appealing to make use of IRS deployment to enhance spatial multiplexing gains of MIMO systems. In terms of IRS deployment in MIMO systems,\cite{shuowen-2022} and \cite{ShiJin-Channel} demonstrated that the deployment of IRSs is able to create favorable channels with controllable rank and favorable condition numbers. For a multi-user setup, \cite{Guangji-2023} rigorously proved that distributed IRS configurations outperform centralized IRS when the total number of IRS elements exceeds a certain threshold. Unfortunately, no existing works have investigated how to configure IRS positions and their corresponding elements in multi-IRS aided MIMO systems to achieve a balance between spatial multiplexing gain and passive beamforming gain, which thus motivates our work. 

\begin{figure}[!t]
    \centering
    \includegraphics[width=0.38\textwidth]{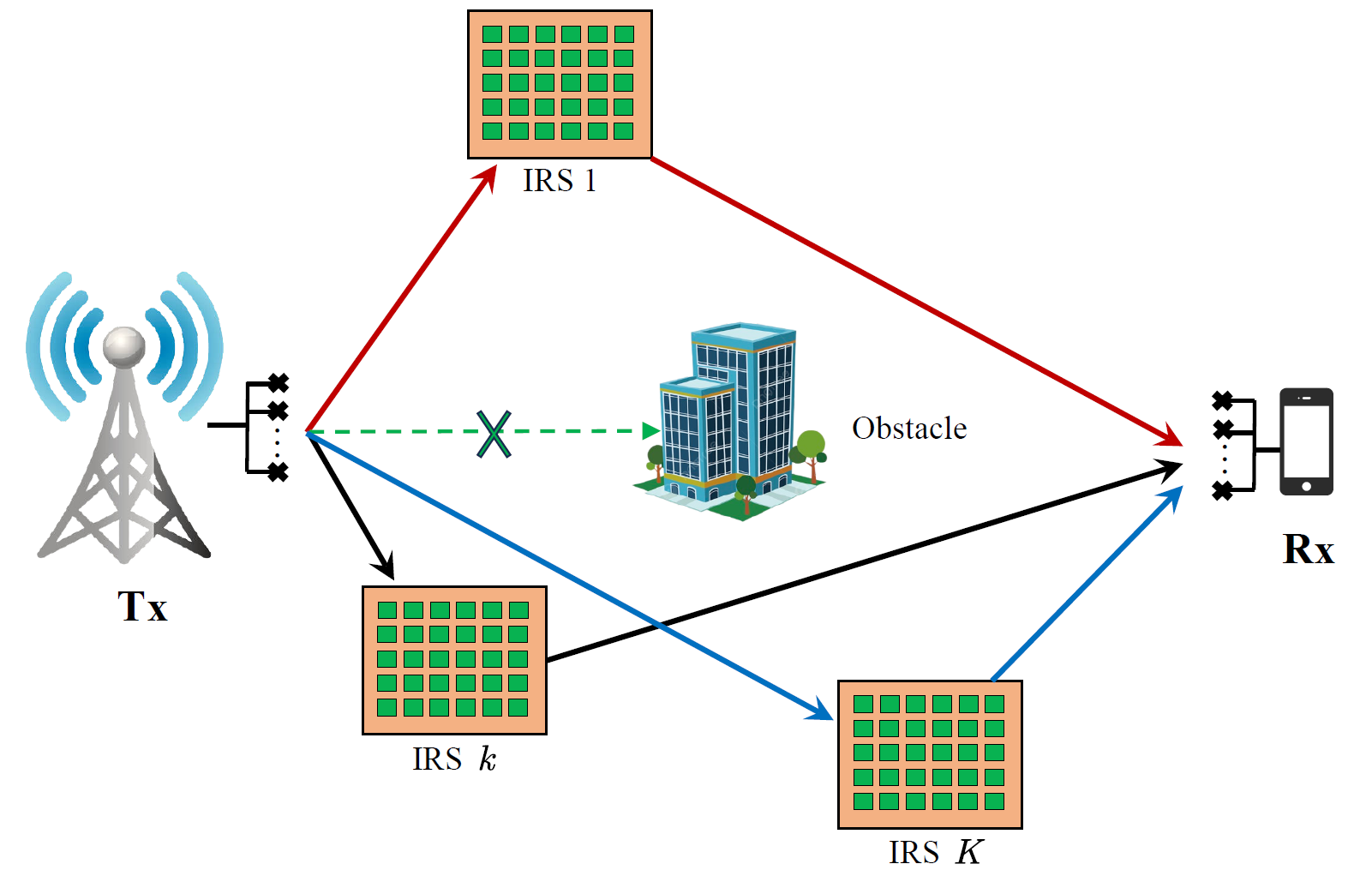}
    \caption{A MIMO wireless communication system aided by $K$ IRSs.}
    \label{fig1}
\end{figure}

In this paper, we focus on a multi-IRS assisted point-to-point MIMO system, as illustrated in Fig. \ref{fig1}. We aim to configure a favorable wireless propagation environment for supporting a multi-stream transmission. To this end, the positions and beamforming of the IRSs have to be designed to maximize the exploitation of the capability of spatial multiplexing while simultaneously minimizing the link path loss. Furthermore, the number of elements and transmit power for each data stream should be carefully allocated to maximize the spectral efficiency (SE). Aiming to address these issues, the main contributions of this work are summarized as follows. First, we propose an IRSs placement scheme to create the orthogonal sub-channels and thereby the decoupled spatial correlation can be fulfilled. Under the orthogonal IRS placement, we unveil that the optimal IRS phase shifts are set to maximize the power gain of each individual sub-channel. Second, we propose an efficient algorithm that jointly optimizes the IRS elements allocation and power to maximize the SE. Moreover, we analytically characterize the scaling orders of the system SE with respect to the number of reflecting elements and power. Finally, numerical results are provided to compare the performance of the multi-IRS aided wireless system with various benchmark systems. Compared to a single large IRS, the potential benefits of multiple IRSs are fully unleashed as the increase of the total number of elements or transmit power. 
%Section II
\section{System Model and Problem Formulation}\label{SecII}
\subsection{System Model}
As shown in Fig. \ref{fig1}, we consider a MIMO wireless communication system assisted by $K$ IRSs, where the Tx and the Rx are equipped with $N_t$ and $N_r$  antennas, respectively. We assume that the direct Tx-Rx link is obstructed. The positions of the Tx and the Rx in a two-dimensional (2D) Cartesian coordinate system are denoted by $\mathbf {u}_{t}\in \mathbb {R}^{2\times 1}$ and $\mathbf {u}_{r}\in \mathbb {R}^{2\times 1}$, respectively. The Tx and the Rx employ uniform linear arrays (ULAs) with element spacings of $d_t$ and $d_r$, respectively. The $k$th IRS is equipped with an $M_k\,(M_{\mathrm{ v,k}}\times M_{\mathrm{ h,k}})$ elements uniform planar array (UPA), where $k\in {\mathcal K}\buildrel \Delta \over =\{1, 2, \ldots, K\}$. The UPA on the $k$th IRS consists of $M_{\mathrm{ v,k}}$ rows and $M_{\mathrm{ h,k}}$ columns, with all spacings equal to $d_s$. The total available number of IRSs is $M$ and thus $\sum_{k=1}^K{M_k}\leqslant M$. The position of the $k$th IRS is denoted by $\mathbf {u}_{k}\in \mathbb {R}^{2\times 1}$.

We denote the equivalent channel from the Tx to the $k$th IRS and from the $k$th IRS to the Rx as $\mathbf {T}_{k}\in \mathbb {C}^{M_{k}\times N_{t}}$ and $\mathbf {R}_{k}\in \mathbb {C}^{N_{r}\times M_{k}}$, respectively. We assume that the distributed IRSs possess line-of-sight (LoS) paths between the Tx and the Rx. The channel from the Tx to the $k$th IRS can be expressed as
\begin{align}\label{channel_Tk_LoS}
\mathbf{T}_k=\rho _{\mathrm{T},k}\mathbf{a}_{\mathrm{S},k}\left( \Phi _{\mathrm{T},k}^{\mathrm{AOA}},\Theta _{\mathrm{T},k}^{\mathrm{AOA}} \right) \mathbf{a}_{N_t}^{H}\left( \Theta _{\mathrm{T},k}^{\mathrm{AOD}} \right) ,
\end{align}
where $\rho _{\mathrm{T},k}$ denotes the complex channel gain of the Tx to the $k$th IRS link. In the array response vector, ${\Phi _{{\mathrm{T}},k}^{\mathrm{AOA}}}=2\pi d_{\mathrm{ S}}\cos {\phi _{{\mathrm{T}},k}^{\mathrm{AOA}}}/\lambda$, ${\Theta _{{\mathrm{T}},k}^{\mathrm{AOA}}}=2\pi d_{\mathrm{ S}}\sin {\phi _{{\mathrm{T}},k}^{\mathrm{AOA}}}\sin {\theta _{{\mathrm{T}},k}^{\mathrm{AOA}}}/\lambda$, and ${\Theta _{{\mathrm{T}},k}^{\mathrm{AOD}}}=2\pi d_{{ t}}\sin {\theta _{{\mathrm{T}},k}^{\mathrm{AOD}}}/\lambda$, where ${\theta _{{\rm{T}},k}^{\rm{AOA}}}$, ${\phi _{{\rm{T}},k}^{\rm{AOA}}}$,  and ${\theta _{{\rm{T}},k}^{\rm{AOD}}}$ denote the horizontal angle of arrival (AoA), the vertical AoA, and the angle of departure (AoD) of the Tx to $k$th IRS link, respectively. Furthermore, ${{\bf{a}}_{N_t}}\left(  \cdot  \right)$ and ${{\bf{a}}_{{\rm{S}},k}}\left(  \cdot  \right)$ represent the array response vectors at the Tx and the $k$th IRS, respectively. Hence, the array response vector of ULA can be unified by 
\begin{align}\label{ULA}
\mathbf{a}_N\left( X \right) =\left[ 1,e^{jX},...,e^{jX\left( N-1 \right)} \right] ^T.
\end{align}
It is worth noting that the array response for UPA can be decomposed into that of ULA as ${{\bf{a}}_{{\rm{S}},k}}\left( {X,Y} \right) = {{\bf{a}}_{{M_{v,k}}}}\left( X \right) \otimes {{\bf{a}}_{{M_{h,k}}}}\left( Y \right)$, where $\otimes$ is the Kronecker product. 

Similar to the Tx to the $k$th IRS link, the channel matrix from the $k$th IRS to the Rx can be expressed as
\begin{align}\label{channel_Rk_LoS}
\mathbf{R}_k=\rho _{\mathrm{R},k}\mathbf{a}_{N_r}\left( \Theta _{\mathrm{R},k}^{\mathrm{AOA}} \right) \mathbf{a}_{\mathrm{S},k}^{H}\left( \Phi _{\mathrm{R},\mathrm{k}}^{\mathrm{AOD}},\Theta _{\mathrm{R},k}^{\mathrm{AOD}} \right),
\end{align}
where $\rho _{\mathrm{R},k}$ denotes the complex channel gain of the $k$th IRS to the Rx link. For notational convenience, in the  sequel we substitute $\mathbf{a}_{\mathrm{S},k,\mathrm{T}}$ and $\mathbf{a}_{\mathrm{S},k,\mathrm{R}}$ for $\mathbf{a}_{\mathrm{S},k}\left( \Phi _{\mathrm{T},k}^{\mathrm{AOA}},\Theta _{\mathrm{T},k}^{\mathrm{AOA}} \right) 
$ and $\mathbf{a}_{\mathrm{S},k}\left( \Phi _{\mathrm{R},\mathrm{k}}^{\mathrm{AOD}},\Theta _{\mathrm{R},k}^{\mathrm{AOD}} \right) 
$, respectively. Besides, we define $\rho _k\triangleq \rho _{\mathrm{T},k}\rho _{\mathrm{R},k}$ and we denote $\mathbf{\Phi }_k=\mathrm{diag}(e^{j\phi _{k,1}},e^{j\phi _{k,2}},\cdots ,e^{j\phi _{k,M_k}})$ as the passive beamforming matrix of the $k$th IRS, where $\phi _{k,m_k}$ is the phase of the $m_k$th element on the $k$th IRS, $m_k\in \mathcal{M} _k\triangleq \left\{ 1,2,…,M_k \right\}$, $k\in \mathcal{K}$, and $\phi _{k,m_k}\in \left[ 0,2\pi \right)$. As such, the effective Tx-IRS-Rx MIMO channel aided by $K$ IRSs is given by $\mathbf{H}=\sum_{k=1}^K{\mathbf{R}_k\mathbf{\Phi }_k\mathbf{T}_k}$.

Let $f\left( \mathbf{\Phi }_k \right) \triangleq \mathbf{a}_{\mathrm{S},k,\mathrm{R}}^{H}\mathbf{\Phi }_k\mathbf{a}_{\mathrm{S},k,\mathrm{T}}$, $\omega _k=\angle (\rho _kf\left( \mathbf{\Phi }_k \right))$, $\mathbf{A}_{\mathrm{T}}=\frac{1}{\sqrt{N_t}}[e^{j\omega _1}\mathbf{a}_{N_t}\left( \Theta _{\mathrm{T},1}^{\mathrm{AOD}} \right) ,...,e^{j\omega _K}\mathbf{a}_{N_t}\left( \Theta _{\mathrm{T},K}^{\mathrm{AOD}} \right) ]
$, and $\mathbf{A}_{\mathrm{R}}=\frac{1}{\sqrt{N_r}}[\mathbf{a}_{N_r}\left( \Theta _{\mathrm{R},1}^{\mathrm{AOA}} \right) ,...,\mathbf{a}_{N_r}\left( \Theta _{\mathrm{R},K}^{\mathrm{AOA}} \right) ]$. Thus, the composite channel consisting of $K$ IRSs can be simplified as
\begin{align}\label{NewChannel}
\!\!\!\mathbf{H}\!=\!\sum_{k=1}^K{\rho _k\mathbf{a}_{N_r}\left( \Theta _{\mathrm{R},k}^{\mathrm{AOA}} \right) f\left( \mathbf{\Phi }_k \right)}\mathbf{a}_{N_t}^{H}\left( \Theta _{\mathrm{T},k}^{\mathrm{AOD}} \right) \!
\!=\!\mathbf{A}_{\mathrm{R}}\mathbf{\Sigma A}_{\mathrm{T}}^{H},
\end{align}
where $\mathbf{\Sigma }=\sqrt{N_tN_r}\mathrm{diag}\left( |\rho _1f\left( \mathbf{\Phi }_1 \right) |,...,|\rho _Kf\left( \mathbf{\Phi }_K \right) | \right) 
$. As such, the received signal $\mathbf {y}\in \mathbb {C}^{N_{r}\times 1}$ at the Rx is given by $\mathbf{y}=\mathbf{Hx}+\mathbf{z}$, where $\mathbf {x}\in \mathbb {C}^{N_{t}\times 1}$ denotes the transmitted signal vector and $\mathbf {z}\sim \mathcal {CN}(0,\sigma ^{2}\mathbf {I}_{N_{r}})$ is the additive white Gaussian noise at the Rx with power $\sigma ^{2}$.

The introduction of IRSs enables the creation of a controllable scattering channel environment, potentially establishing favorable conditions for multi-stream transmission. By properly positioning IRSs so that $\mathbf{A}_{\mathrm{T}}$ and $\mathbf{A}_{\mathrm{R}}$ are orthogonal matrices (i.e., $\mathbf{A}_{\mathrm{R}}^{H}\mathbf{A}_{\mathrm{R}}=\mathbf{I}$ and $\mathbf{A}_{\mathrm{T}}^{H}\mathbf{A}_{\mathrm{T}}=\mathbf{I}$), the equation \eqref{NewChannel} naturally admits a form of the singular value decomposition (SVD). In this scenario, the structures of the transmitter's precoding and the receiver's combiner can be explicitly derived as $\mathbf{A}_{\mathrm{T}}$ and $\mathbf{A}_{\mathrm{R}}^{H}$, respectively. This configuration allows for the minimization of inter-stream interference while simplifying the transceiver design. 
\subsection{Problem Formulation}
We aim to maximize the SE of the multi-IRS aided system by optimizing the IRS beamforming, IRS placement, elements allocation, and transmit covariance matrix. The corresponding optimization problem is formulated as
\begin{align}
     \mathop {\mathrm {max} }\limits _{ { \mathop{\{u_{k}\}, \mathbf{Q}}\limits^{ \{\phi _{k,m_k}\},\{M_{k}\}} } } &\log _2\det \left( \mathbf{I}_{N_r}+\frac{1}{\sigma ^2}\mathbf{HQH}^H \right) \label{P1}\\
    \text{s.t.} \ 
    &\phi _{k,m_k}\in \left[ 0,2\pi \right),m_k\in \mathcal{M}_k,k\in \mathcal{K},\tag{\ref{P1}{a}} \label{opt-cstrt-a} \\
    &\mathrm{tr(}\mathbf{Q})\le P,\mathbf{Q}\succeq \mathbf {0} ,\tag{\ref{P1}{b}} \label{opt-cstrt-b} \\
&\mathbf{A}_{\mathrm{R}}^{H}\mathbf{A}_{\mathrm{R}}=\mathbf{I},\mathbf{A}_{\mathrm{T}}^{H}\mathbf{A}_{\mathrm{T}}=\mathbf{I}, \tag{\ref{P1}{c}}\label{opt-cstrt-c} \\
    &\sum\nolimits_{k=1}^K{M_k}\leqslant M,M_k\in \mathbb{Z} _{\geqslant 0},k\in \mathcal{K}, \tag{\ref{P1}{d}} \label{Element Constraints}
\end{align}
where $\mathbf {Q}=\mathbb {E}\{\mathbf {x}\mathbf {x}^{H}\}$ denotes the transmit covariance matrix and $P$ denotes the maximum transmit power of the Tx.

\section{Proposed Solution} \label{Sec::opt}

Problem \eqref{P1} is non-convex due to its non-concave objective function, IRS position constraints, and integer elements allocation constraints. Generally, there are no standard methods to solve it optimally. To address this issue, we first propose an efficient solution to determine the IRS placement. By exploiting the particular structure provided by the IRS placement, we derive the optimal IRS beamforming in a closed form expression. Subsequently, the IRS elements allocation and the power allocation are jointly optimized.

\subsection{IRS Placement and Beamforming Design} \label{Subsec::opt-solution}
First, to satisfy the constraint \eqref{opt-cstrt-c}, IRSs should be positioned along the discrete fourier transform (DFT) directions of both the Tx and the Rx. This arrangement satisfies the requirement for the AoA and AoD discretization:
\begin{align}\label{Deployment Condi} 
\begin{cases}
	\Theta _{\mathrm{T},k}^{\mathrm{AOD}}\in \mathcal{A} _1\triangleq \left\{ \frac{2\pi i}{N_t}-\pi \right\} _{i=1}^{N_t},k\in \mathcal{K} ,\\
	\Theta _{\mathrm{R},k}^{\mathrm{AOA}}\in \mathcal{A} _2\triangleq \left\{ \frac{2\pi i}{N_r}-\pi \right\} _{i=1}^{N_r},k\in \mathcal{K} ,\\
	\Theta _{\mathrm{T},i}^{\mathrm{AOD}}\ne \Theta _{\mathrm{T},j}^{\mathrm{AOD}},\Theta _{\mathrm{R},i}^{\mathrm{AOA}}\ne \Theta _{\mathrm{R},j}^{\mathrm{AOA}},i,j\in \mathcal{K} ,i\ne j.\\
\end{cases}
\end{align}

When the positions of the Tx and the Rx as well as $(\Theta _{\mathrm{T},k}^{\mathrm{AOD}},\Theta _{\mathrm{R},k}^{\mathrm{AOA}})$ are given, we can determine the corresponding IRS position $\mathbf {u}_{k}$. However, obtaining the optimal IRS positions requires an exhaustive enumeration of all combinations due to \eqref{Deployment Condi}, which results in high computation complexity. To address this problem, we employ a greedy algorithm that searches for the position with the maximum channel gain at each iteration. Besides, we define $\mathcal{P} _1\triangleq \left\{ \mathcal{S} _1,\mathcal{S} _2,…,\mathcal{S} _L \right\} $ that collects all the possible IRS candidate positions and their corresponding Tx-IRS-Rx channel gain, where $L$ denotes the maximum number of different sets in $\mathcal{P}_1$, $\mathcal{S} _l\triangleq \left( \theta _l,\varphi _l,\tau _l \right)$ denotes the IRS position where the AoD at the Tx is $\theta_l$ and the AoA at the Rx is $\varphi_l$ with the corresponding Tx-IRS-Rx channel gain $\tau _l$, $\theta _l\in \mathcal{A} _1$, $\varphi _l\in \mathcal{A} _2$, $l\in \mathcal{L} \triangleq \left\{ 1,2,…,L \right\}$, and $\mathcal{S}_i\ne \mathcal{S}_j,i\ne j,i,j\in \mathcal{L}$. Then, the process of the greedy search can be described by
\begin{align} \label{NewGreed}
\begin{aligned}
&\left( \Theta _{\mathrm{T},k+1}^{\mathrm{AOD}},\Theta _{\mathrm{R},k+1}^{\mathrm{AOA}},\rho _{k+1} \right) =\mathop {\mathrm{arg}\max}_{\mathcal{S} _l\in \mathcal{P} _{k+1}}\tau _l\\
&\text{s.t.}\ \begin{aligned}[t]
    &\mathcal{P}_{k+1} = \mathcal{P}_k \backslash \mathcal{B}_k, k\geqslant 1, \\
    &\mathcal{B}_k = \left\{ \mathcal{S}_i \in \mathcal{P}_k \,|\, \theta_i = \Theta_{\mathrm{T},k}^{\mathrm{AOD}} \text{ or } \varphi_i = \Theta_{\mathrm{R},k}^{\mathrm{AOA}} \right\}.
\end{aligned}
\end{aligned}
\end{align}
Therefore, we execute this greedy search process from $k=0$ to $K-1$ and successfully obtain the positions for $K$ IRSs.
% where $\mathcal{P} _1=\mathcal{P}$, $\mathcal{P} _k\triangleq \left\{ s_{k,1},s_{k,2},…,s_{k,L_k} \right\} $, $s_{k,l_k}\triangleq \left( \theta _{l_k},\varphi _{l_k},D_{l_k} \right) \in \mathcal{P} _k$, and $l_k\in \mathcal{L} _k\triangleq \left\{ 1,2,…,L_k \right\}$. In order to ensure \eqref{Deployment Condi}, $\mathcal{P} _k$ is derived from $\mathcal{P} _{k-1}$ by removing $s_{k-1,l_{k-1}}$ containing $\theta _{l_{k-1}}=\Theta _{\mathrm{T},k-1}^{\mathrm{AOD}}$ or $\varphi _{l_{k-1}}=\Theta _{\mathrm{R},k-1}^{\mathrm{AOA}}$ when $k\geqslant 2$. Therefore, we execute this greedy search process from $k=1$ to $K$ and successfully obtain the positions for $K$ IRSs.

By employing the orthogonal placement, the $k$th singular value of $\mathbf {H}$ is the non-zero singular value of $\mathbf{R}_k\mathbf{\Phi }_k\mathbf{T}_k$. And $\mathbf{Q}$ in \eqref{P1} can be derived as $\mathbf{Q}=\mathbf{A}_{\mathrm{T}}\mathrm{diag}\left( p_{1},p_{2},…,p_{K}\right) \mathbf{A}_{\mathrm{T}}^{H}$, where $p_{k}$ is the transmit power allocated to the $k$th sub-channel that will be optimized in subsection B. Thus, the SE of the $K$ IRS system can be rewritten as
\begin{align}
R=\sum\nolimits_{k=1}^K{\log _2}\left( 1+p_k\chi _k\left( f\left( \mathbf{\Phi }_k \right) \right) ^2 \right) ,
\end{align}
where $\chi _k=|\rho _k|^2/\sigma ^2$. Due to the fact that maximizing the SE can be achieved by independently maximizing each $f( \mathbf{\Phi }_k)$, we employ the following passive beamforming structure to optimize each $f( \mathbf{\Phi }_k)$:
\begin{align}\label{Beamforming}
\phi _{k,m_k}=\mathrm{arg}\left( (\mathbf{t}_k)_{m_k}^{\ast}(\mathbf{r}_k)_{m_k}^{\ast} \right),m_k\in \mathcal{M}_k,k\in \mathcal{K},
\end{align}
where $(\mathbf{t}_{k})_{m_k}$ is the $m_k$th element of $\mathbf{a}_{N_t}^{H}\left( \Theta _{\mathrm{T},k}^{\mathrm{AOD}} \right) 
$ and $(\mathbf{r}_{k})_{m_k}$ is the $m_k$th element of $\mathbf{a}_{N_r}\left( \Theta _{\mathrm{R},k}^{\mathrm{AOA}} \right) 
$. By employing the above configuration, $f\left( \mathbf{\Phi }_k \right)$ can be maximized. Under the optimal $\mathbf{\Phi }_{k}^{*}$, we have $f\left( \mathbf{\Phi }_{k}^{*} \right)=M_k$. 

\subsection{Joint IRS Elements Allocation and Power Optimization} 

To better illustrate the spatial multiplexing gain offered by multiple IRSs, we unveil the condition that double-IRS outperforms single-IRS in the following proposition. Besides, we define $R_1=\log _2\left( 1+\chi PM^2 \right)$ is the maximum achievable rate with a single IRS comprising $M$ elements and $R_2=\underset{p_1,p_2}{\max}\left( \log _2\left( 1+p_1\eta _1 \right) +\log _2\left( 1+p_2\eta _2 \right) \right) $ with $p_1+p_2=P$ and $\eta_k =\chi_k M^2/4$ is the maximum achievable rate with two IRSs, each comprising $M_1=M_2=M/2$ elements under orthogonal placement conditions.
\begin{pos}
When $\chi _1=\chi _2=\chi$, we have $R_2\geqslant R_1$ if 
\begin{align} \label{laji}
    \chi PM^2\geqslant 48.
\end{align}

\begin{proof}
According to the water-filling power allocation\cite{CVX}, for $R_2$ the optimal $p_{k}^{*}$ can be derived as $p_{k}^{*}=\max \left( u-1/\eta_k ,0 \right),k\in \left\{ 1,2 \right\}$. Since $\eta_1=\eta_2$, we further have $p_{k}^{*}=P/2$ and $R_2=2\log _2\left( 1+\chi PM^2/8 \right) $. To satisfy $R_2\geqslant R_1$, we can obtain \eqref{laji}, which completes the proof.
\end{proof}
\end{pos}

Proposition 1 theoretically illustrates that, under specific conditions and with appropriate allocation strategies, the performance of a double IRS system can surpass that of a single IRS configuration. Furthermore, we aim to fully exploit the spatial multiplexing potential of multi-IRS architectures through appropriate resource allocation strategies, which motivates us to investigate the joint power and elements allocation algorithm.
Next, we jointly optimize the IRS elements and power allocation under the obtained IRS placement and beamforming. To this end, problem \eqref{P1} is reduced to
\begin{align} 
   (\mathrm{P1}):\
    \underset{\left\{ p_k \right\} ,\left\{ M_k \right\}}{\max}&\sum\nolimits_{k=1}^K{\log _2}\left( 1+p_kM_{k}^{2}\chi _k \right)  \label{P11}\\
    \text{s.t.} \ 
    &\sum\nolimits_{k=1}^K{p_k}\leqslant P, p_k\geqslant 0,k\in \mathcal{K},\tag{\ref{P11}{a}} \label{Power Constraints} \\
    % &p_k\geqslant 0,k\in \mathcal{K},\tag{\ref{P11}{b}} \label{Power Positivity} \\
    &\eqref{Element Constraints}. \nonumber
\end{align}

First, to address the integer constraint in \eqref{Element Constraints}, we relax the discrete value $M_k$  to its continuous counterpart $\tilde{M}_k$. Consequently, the objective function of problem \eqref{P11} can be relaxed to
\begin{align}\label{Continuousize  OF}
\sum\nolimits_{k=1}^K{\log _2}\left( 1+p_k\tilde{M}_{k}^{2}\chi _k \right).
\end{align}
However, \eqref{Continuousize  OF} is non-convex due to the coupling of $p_k$ and $\tilde{M}_{k}$. To tackle this difficulty, we introduce auxiliary variables $l_k$. Problem $(\mathrm{P1})$ can be reformulated as the following optimization problem: 
\begin{align} \underset{\left\{ p_k \right\} ,\left\{ \tilde{M}_k \right\} ,\left\{ l_k \right\}}{\max}&\sum\nolimits_{k=1}^K{\log _2}\left( 1+l_k \right) 
    \label{P3}\\
    \text{s.t.} \ 
    &l_k\leqslant \chi _kp_k\tilde{M}_{k}^{2},k\in \mathcal{K}, \tag{\ref{P3}{a}}\label{Auxiliary Constraints 1} \\
    &\sum\nolimits_{k=1}^K{\tilde{M}_k}\leqslant M, \tilde{M}_k\geqslant 0,k\in \mathcal{K},\tag{\ref{P3}{b}} \label{Continuous Element} \\
    &\eqref{Power Constraints}. \nonumber
\end{align}

Problem \eqref{P3} is still intractable hard to solve due to \eqref{Auxiliary Constraints 1}. To address the non-convexity of constraint \eqref{Auxiliary Constraints 1}, we employ the successive convex approximation (SCA) technique with appropriate variable substitutions. Thus, we introduce auxiliary variables $x_k$ and $y_k$, $k\in \mathcal{K}$. Consequently, the constraints in \eqref{Auxiliary Constraints 1} can be reformulated as follows:
\begin{align}
\label{SCA1} &l_k\leqslant \chi _ke^{x_k+y_k}, e^{y_k}\leqslant \tilde{M}_{k}^{2},k\in \mathcal{K},
\\
\label{SCA3}&e^{x_k}\leqslant p_k,k\in \mathcal{K}.
\end{align}
Although constraints \eqref{SCA1} exhibit non-convex forms, their right-hand sides, specifically $\chi _ke^{x_k+y_k}$ and $\tilde{M}_{k}^{2}$, are convex functions with respect to their respective variables. This motivates us to apply the first-order Taylor expansion to linearize them as convex constraints given by
\begin{align}
&l_k\leqslant \chi _ke^{\hat{x}_k+\hat{y}_k}\left( 1+x_k+y_k-\hat{x}_k-\hat{y}_k \right),k\in \mathcal{K},\label{NewSCA1}
\\
&e^{y_k}\leqslant 2\tilde{M}_k\hat{M}_k-\hat{M}_{k}^{2},k\in \mathcal{K},\label{NewSCA2}
\end{align}
where $\hat{x}_k$, $\hat{y}_k$, and $\hat{M}_k$ are the given local points of $x_k$ ,$y_k$, and $\tilde{M}_{k}$, respectively. 

As a result, problem \eqref{P3} is approximated as
\begin{align} 
(\mathrm{P2}):\
\underset{\left\{ p_k \right\} ,\left\{ \tilde{M}_k\right\} ,\left\{ l_k \right\}}{\max}&\sum\nolimits_{k=1}^K{\log _2}\left( 1+l_k \right) 
    \label{P4}\\
    \text{s.t.} \ 
    &\eqref{Power Constraints},\eqref{Continuous Element},\eqref{SCA3},\eqref{NewSCA1},\eqref{NewSCA2}.\nonumber
\end{align}

Problem $(\mathrm{P2})$ is a convex optimization problem, which can be solved optimally in an iterative manner by using standard solvers like CVX until convergence is achieved. As the objective value increases with each iteration and is bounded by a finite value, the solution of the original problem $(\mathrm{P1})$ is ensured to reach convergence. The integer number of reflecting elements can be determined by rounding the continuous solutions to problem $(\mathrm{P2})$. Moreover, the complexity of the greedy search algorithm in \eqref{NewGreed} is smaller than $\mathcal{O} \left( KN_tN_r \right)$ and the complexity of solving problem $(\mathrm{P2})$ via SCA is $\mathcal{O} \left( JK^{3.5} \right)$ with $J$ denoting the number of iteration while that of obtaining the optimal IRS beamforming is negligible due to the closed-form solution given in \eqref{Beamforming}.

\section{SE Scaling Order Analysis}

In this section, we characterize the SE scaling order with respect to $M$ and $P$ for the multi-IRS aided wireless system. Based on the analysis in Section III, we first derive the elements allocation and power allocation scheme for the multi-IRS as $M$ approaches infinity.
\begin{pos}
Under orthogonal placement conditions, as ${M} \to \infty$ and we assume that $M$ is a multiple of $K$, the asymptotically optimal solution of problem $(\mathrm{P1})$, denoted by $\left\{ {p_{k}^*,M_k^*} \right\}$, is derived as
\begin{align}\label{optimal_solution2}
M_{k}^{*}=\frac{M}{K}
,p_{k}^{*}=\frac{P}{K}
, k\in \mathcal{K} ,
\end{align}
\begin{proof}
Note that it can be easily proved that the solution obtained when not using all IRS elements is always suboptimal as ${M} \to \infty$. Under any given $M_k$, the optimal solution for $p_{k}$ is $p_k=\max \left( u-1/\left( \chi _kM_k^{2} \right) ,0 \right) $ according to the water-filling algorithm. As $M\to \infty $, $p_k=u-1/\left( \chi _kM_k^{2} \right) $ holds naturally. Thus, the objective function of problem \eqref{P11} can be simplified as
\begin{align}
\underset{M\rightarrow \infty}{\lim}\sum_{k=1}^K{\log _2}\left( 1+p_kM_{k}^{2}\chi _k \right) =\sum_{k=1}^K{\log _2}\left( p_kM_{k}^{2}\chi _k \right).
\end{align}

According to the Cauchy-Schwarz inequality, we have 
\begin{align}
\varGamma _k\left( \prod_{k=1}^K{M_k} \right) ^2\leqslant \varGamma _k\left( \sum_{k=1}^K{M_k}/K \right) ^{2K},
\end{align}
where $\varGamma _k=\prod_{k=1}^K{p_k\chi _k}$, $\sum_{k=1}^K{M_k}=M$ and the condition for equality in this case is ${M}_{k}^{*}={M}/K, k \in {\cal K}$. Accordingly, the optimal power allocation in this case is $P/K, \forall k \in {\cal K}$ .
% and its resultant objective value is 
% \begin{align}
%  C=\sum_{k=1}^K{\log _2}\left( 1+\frac{PM^{2}}{K^3}\chi _k \right).
% \end{align}
Thus, we complete the proof.
\end{proof}
\end{pos}

Proposition 2 reveals that, given a large number of elements, equal elements/power allocation can achieve near-optimal performance. This is because deploying a large number of IRS elements significantly enhances the magnitude of the singular values of the link, thereby artificially creating an equivalent "high-SNR" condition. 
\begin{pos}
Under orthogonal placement conditions, as ${M} \to \infty$, the system SE increases with ${M}$ according to
\begin{align}\label{propo1}
\lim_{M\rightarrow \infty} \frac{R}{\log _2M}=2K.
\end{align}
\begin{proof}
According to Proposition 2, when ${M} \to \infty$  we have
\begin{align} 
\lim_{M\rightarrow \infty} R=\lim_{M\rightarrow \infty} \sum_{k=1}^K{\left( \log _2\left( \varGamma _k/K \right) +2\log _2\left( M \right) \right)}.
\end{align}

As such, we have \eqref{propo1}, which completes the proof.
\end{proof}
\end{pos}
\begin{pos}
Under orthogonal placement conditions, as ${P} \to \infty$, the system SE increases with $P$ according to
\begin{align}\label{propo2}
\lim_{M\rightarrow \infty} \frac{R}{\log _2P}=K.
\end{align}

\begin{proof}
Similar to Proposition 3, when ${P} \to \infty$  we have
\begin{align}
\lim_{P\rightarrow \infty} R=\lim_{P\rightarrow \infty} \sum_{k=1}^K{\left( \log _2\left( M_{k}^{2}\chi _k/K \right) +\log _2\left( P \right) \right)}.
\end{align}

As such, we have \eqref{propo2}, which completes the proof. 
\end{proof}
\end{pos}

Proposition 3 and Proposition 4 demonstrate that compared to the single IRS system ($K=1$), a $K$-fold gain is realized in the SE scaling order of either $M$ or $P$, which implies that we can potentially harvest a spatial multiplexing gain of a factor of $K$ from the system by deploying $K$ IRSs. 

\vspace{-2mm}
\section{Numerical Results}
In this section, numerical results are provided to compare the performance of multiple IRSs and single IRS configurations, as well as to draw useful insights. The locations of the Tx and the Rx are set at (0, 0) and (85m, 0) respectively. We deploy a ULA at both the Tx and the Rx, aligning the base directions of the antenna arrays at the Tx and the Rx as $\mathbf{v}_t=\mathbf{v}_r=[0,-1]^T$. Other system parameters are configured as follows: $N_t=8$, $N_r=4$, $\lambda =0.15$ m, $H=5$ m, $d_t=d_r=\lambda /2=0.075$ m, and $\sigma^{2}=-80$ dBm.

\begin{figure}[t]
	\centering
	\includegraphics[width=2.3in]{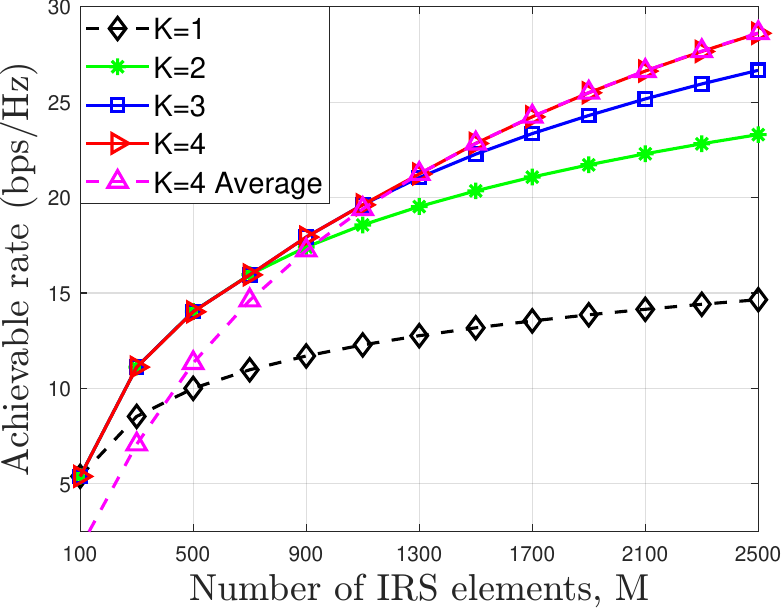}
	\vspace{-3pt}
	\caption{Achievable rate versus $M$ under orthogonal placement.}
	\label{fig2}
	\vspace{-10pt}
\end{figure}

\begin{figure}[t]
	\centering
	\includegraphics[width=2.3in]{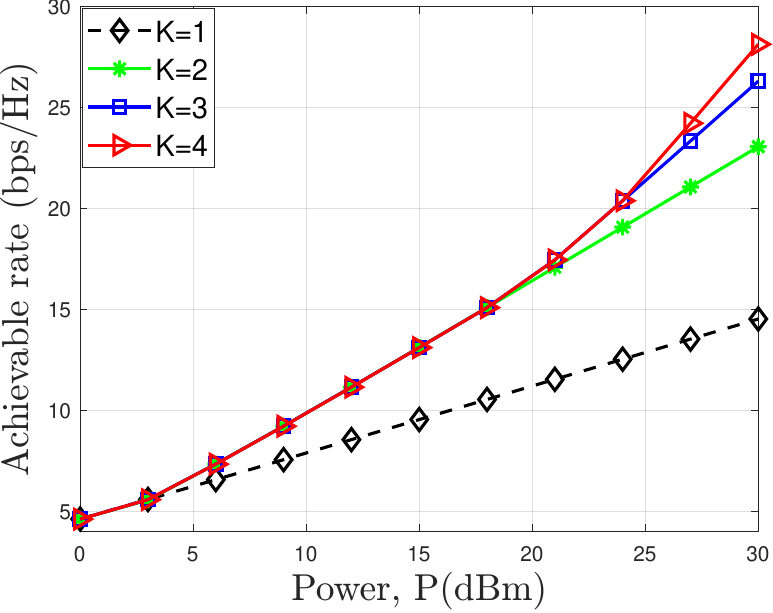}
	\vspace{-3pt}
	\caption{Achievable rate versus $P$ under orthogonal placement.}
	\label{fig3}
	\vspace{-10pt}
\end{figure}

Fig. \ref{fig2} and Fig. \ref{fig3} compare the performance of the multi-IRS and the single-IRS aided systems under orthogonal placement. For the multi-IRS aided MIMO system, we employ our proposed algorithm to maximize the SE. Fig. \ref{fig2} and Fig. \ref{fig3} compare the achievable rates for different numbers of $K$ under the conditions of $P = 30$ dBm and $M$ = 2400, respectively. "$K$=4 Average" strategy refers to the scenario where the power and elements are equally allocated. As shown in Fig. \ref{fig2} and Fig. \ref{fig3}, the multi-IRS aided system significantly outperforms its single-IRS counterpart as the power and the total number of IRS elements increase. However, the single-IRS aided system remains optimal under the constraints of lower power and fewer total IRS elements.

\begin{figure}[t]
	\centering
	\includegraphics[width=2.3in]{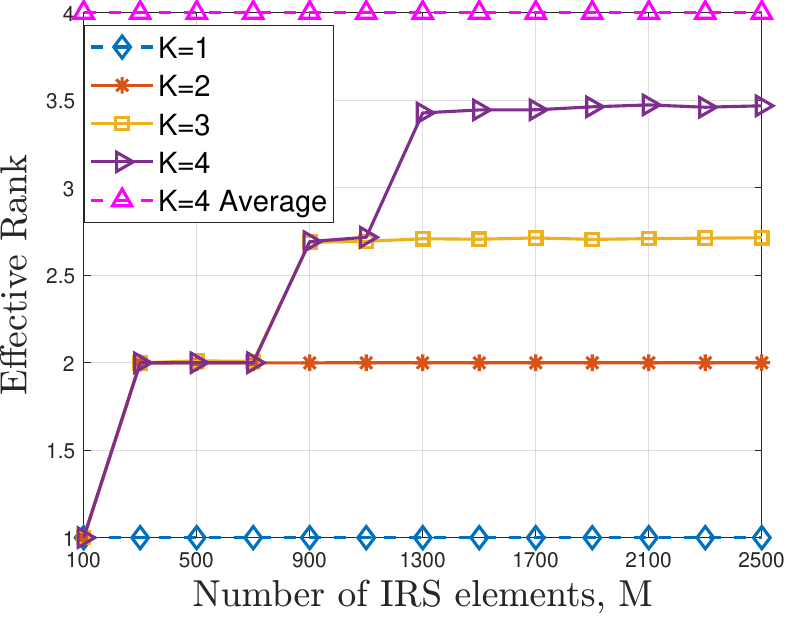}
	\vspace{-3pt}
	\caption{Effective rank versus $M$.}
	\label{fig4}
	\vspace{-10pt}
\end{figure}

In Fig. \ref{fig4}, we further investigate the impact of IRS elements number on the overall system performance. We employ the number of the effective rank, as introduced in \cite{Rank-2007}, which is defined as $\mathrm{Erank}\left( \mathbf{H} \right) =\exp \left( -\sum_k{\bar{\delta}_k\ln \bar{\delta}_k} \right)$,
where $\bar{\delta}_k=\sqrt{\delta _k}/\sum_i{\sqrt{\delta _i}}
$ and $\delta _k$ is the $k$th singular value of $\mathbf{H}$. Fig. \ref{fig4} illustrates the variation of effective rank with the total number of IRS elements when $P=30$ dBm. It is clear that as the number of IRS elements increases, multiple IRSs offer greater potential for rank enhancement compared to a single IRS. The rank improvement reflects the system's spatial multiplexing capability. In particular, enhanced spatial multiplexing allows multiple IRSs to achieve higher gains when the total number of IRS elements is large.
\begin{figure}[t]
	\centering
	\includegraphics[width=2.3in]{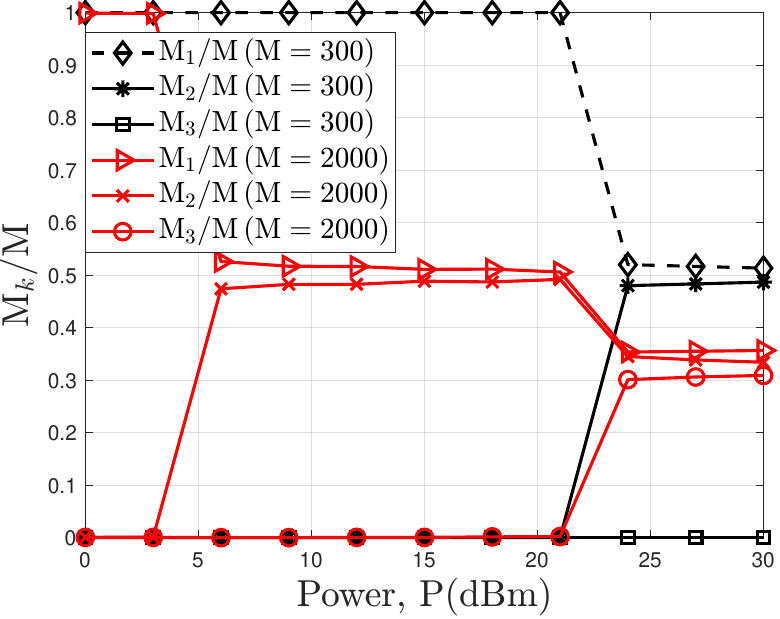}
	\vspace{-3pt}
	\caption{Elements allocation versus $P$ when $K=3$.}
	\label{fig5}
	\vspace{-10pt}
\end{figure}

% \begin{figure}[t]
% 	\centering
% 	\subfigure[Achievable rate versus $M$ under orthogonal placement.]{
% 		\label{fig2a}
% 		\includegraphics[width=4cm]{RElement.png}
% 	}
% 	\subfigure[Achievable rate versus $P$ under orthogonal placement.]{
% 		\label{fig2b}
% 		\includegraphics[width=4cm]{Rpower.png}
% 	}
% 	% \subfigure[The S\&C performance bounds.]{
% 	% 	\label{simufig3}
%  %        \vspace{-1mm}
% 	% 	\includegraphics[width=5.5cm]{fig3-temp-2.eps}
% 	% }
% 	\caption {Performance comparison between our proposed $K$-IRS system and single IRS system. }
% 	\label{fig2}
% \end{figure}

% \begin{figure}[t]
% 	\centering
% 	\subfigure[Effective rank versus $M$.]{
% 		\label{fig3a}
% 		\includegraphics[width=4cm]{Erank.png}

% 	}
% 	\subfigure[Element allocation versus $P$ when $K=3$.]{
% 		\label{fig3b}
% 		\includegraphics[width=4cm]{ElementAllo.png}
              
% 	}
% 	% \subfigure[The S\&C performance bounds.]{
% 	% 	\label{simufig3}
%  %        \vspace{-1mm}
% 	% 	\includegraphics[width=5.5cm]{fig3-temp-2.eps}
% 	% }
% 	\caption {The influence of element number $M$ on the channel rank and the allocation of elements.}
% 	\label{fig3}
% \end{figure}  

Fig. \ref{fig5} provides a more detailed view of the IRS elements allocation process for different total numbers of elements $M$ when $K=3$. In Fig. \ref{fig5}, $M_1$, $M_2$, and $M_3$ denote the number of elements allocated to three IRSs, where $M_1\geqslant M_2\geqslant M_3$ and $\sum_{k=1}^3{M_k}=M$. As the power increases, the IRS with  the largest number of elements will gradually allocate its elements to other IRSs and their elements distribution across individual IRSs tends towards equality, corresponding to our proposition in Section IV.
\vspace{-2mm}
\section{Conclusion}
In this work, we investigate the problem of the IRS placement and resource allocation in MIMO communication systems. We propose an orthogonal placement scheme to maximize the spatial multiplexing gain, upon which we optimize IRS beamforming as well as elements and power allocation to maximize the SE. Moreover, we analytically characterize the system's SE scaling orders with respect to the number of reflecting elements and power. Our numerical results demonstrate that, when the total number of IRS elements or the power exceeds a certain threshold, multi-IRS systems significantly outperform single-IRS systems and equal distribution of elements and power across multiple IRSs is shown to be asymptotically optimal.

\bibliographystyle{IEEEtran}
\bibliography{IEEEabrv,ref}

\end{document}